\documentclass[11pt]{article}
\usepackage[a4paper, lmargin=1in, rmargin=1in, bmargin=1in, tmargin=1in, headheight=10pt]{geometry}
\usepackage[T1]{fontenc}
\usepackage{authblk}

\usepackage{sectsty}
\sectionfont{\normalsize}
\subsectionfont{\normalsize}
\subsubsectionfont{\normalsize}
\paragraphfont{\normalsize}

\usepackage{caption} 
\usepackage[compact]{titlesec}
\titlespacing*{\section}{0pt}{*2.25}{*1.75}
\titlespacing*{\subsection}{0pt}{*2.0}{*1.55}
\titlespacing*{\subsubsection}{0pt}{*1.75}{*1.35}
\usepackage[utf8]{inputenc} 
\usepackage[T1]{fontenc}    
\usepackage{url}            
\usepackage{booktabs}       
\usepackage{amsfonts}       
\usepackage{nicefrac}       
\usepackage{microtype}      
\usepackage{xcolor}         
\usepackage{xr}

\usepackage{amsmath}
\usepackage{graphicx,subcaption}
\usepackage{algorithm}
\usepackage{algpseudocode}
\usepackage{tikz}

\title{Computational design of target-specific linear peptide binders with TransformerBeta}

\author[1,$\dag$,*]{\small Haowen Zhao}
\author[2]{\small Francesco A. Aprile}
\author[1,*]{\small Barbara Bravi}
\affil[1]{Department of Mathematics, Imperial College  London, London SW7 2AZ, United Kingdom}
\affil[2]{Department of Chemistry and Institute of Chemical Biology, Molecular Sciences Research Hub, Imperial College London, London W12 0BZ, United Kingdom}
\affil[$\dag$]{Current affiliation: Department of Chemistry, University of Cambridge, Cambridge CB2 1EW, United Kingdom}
\affil[*]{For correspondence: hz362@cam.ac.uk, b.bravi21@imperial.ac.uk}

\date{}

\begin{document}
\maketitle
\vspace{-3.5em}

\begin{abstract}
\noindent The computational prediction and design of peptide binders targeting specific linear epitopes is crucial in biological and biomedical research, yet it remains challenging due to their highly dynamic nature and the scarcity of experimentally solved binding data. To address this problem, we built an unprecedentedly large-scale library of peptide pairs within stable secondary structures (beta sheets), leveraging newly available AlphaFold predicted structures. We then developed a machine learning method based on the Transformer architecture for the design of specific linear binders, in analogy to a language translation task. Our method, TransformerBeta, accurately predicts specific beta strand interactions and samples sequences with beta-sheet-like molecular properties, while capturing interpretable physico-chemical interaction patterns. As such, it can propose specific candidate binders targeting linear epitope for experimental validation to inform protein design.
\end{abstract}

\section{Introduction}

Peptides have emerged as important tools for fundamental and applied research in protein science and therapeutics. Linear peptide-like binding interfaces are found in many protein-protein interactions involved in cell signalling and regulation \cite{van2014short, wright2009linking, cumberworth2013promiscuity}. Targeting these regions is of great biomedical interest in the treatment of numerous human pathologies, including neurodegenerative diseases \cite{schweers1994structural, weinreb1996nacp, uversky2003protein} and cancer \cite{uyar2014proteome}. Peptides can bind to epitopes with high affinity and specificity, while exhibiting lower immunogenicity and more cost-effective production than large biologics like proteins and antibodies \cite{sharma2022peptide, wang2022therapeutic, davda2019immunogenicity}. 

The design of high-affinity linear peptides can be pivotal for epitope-specific antibody design \cite{sormanni2015rational}. Fragment-based approaches, which design these linear peptides purely by joining interacting protein-protein fragments and grafting them onto antibody scaffolds, have achieved low-nanomolar affinity binders without \textit{in vitro} affinity maturation \cite{sormanni2015rational}. However, the applicability of fragment-based methods to new or less well-characterized targets is limited, as the epitope must already exist in the database, or interacting partners for its short subfragments must be identified and joined using custom rules. Machine learning-based generative models have also shown remarkable potential in designing binding peptides \cite{li2024full, watson2023novo, dauparas2022robust} and antibodies \cite{boom2023score, bennett2024atomically, akbar2022silico, saka2021antibody, leem2022deciphering}. The state-of-the-art method, RFdiffusion, has achieved considerable success in designing medium-sized binding peptides, as validated by a large array of experiments \cite{watson2023novo}. However, RFdiffusion does not focus on designing short linear peptides, and there is room for improvement in the success rate at directly designing binding antibodies \cite{bennett2024atomically}.

Motivated by recent advances in natural language processing, in this paper we propose that training generative language models on protein fragments to learn their complex dependencies could be an effective tool for modelling linear epitope-specific peptide binding, and thus for guiding the rational design of peptide binders and antibodies. These models, first designed for machine translation, have demonstrated broad applicability in generating sequences across many domains \cite{chen2021evaluating, borsos2023audiolm, bagal2021molgpt} and in various protein-related tasks, including identifying epitopes \cite{clifford2022bepipred, chu2022transformer}, learning interaction motifs at the paratope-epitope interface \cite{akbar2021compact}, designing functional proteins \cite{ferruz2022protgpt2, madani2020progen} and peptides with desired biological activities \cite{grisoni2018designing,capecchi2021machine, wang2021deep, caceres2020deep,wu2020signal}.

A major challenge in machine learning-based peptide design is data acquisition, particularly in obtaining a training dataset representative of the complex interaction motif space. The release of the Alphafold Protein Structure Database (Alphafold DB) \cite{varadi2022alphafold}, containing 214 million structures predicted by AlphaFold2 \cite{jumper2021highly}, has provided an unprecedented, large-scale new set of protein structures that can serve many research purposes, from protein design \cite{bryant2022evobind, goverde2023novo} and characterization \cite{yu2022cryo, dauparas2022robust} to training dataset augmentation \cite{hsu2022learning, ruffolo2022fast, hoie2024discotope}. 

In this work, we created AlphaFold 2 Beta Strand Database, hosting 488 million high-quality beta strand interaction motifs collected from Alphafold DB for training data augmentation, since large datasets are crucial to train language models. We then trained a Transformer-based model, that we refer to as TransformerBeta, to predict probabilistic scores of peptide binding to linear epitopes. Based on these scores, TransformerBeta efficiently generates putative peptide binders, with the associated scores useful in candidate selection. We demonstrated that the designed peptides are highly similar to natural interaction motifs, both statistically and physico-chemically, and that the embedding of our model captures biologically meaningful representations. 

\section{Results}

\subsection{AlphaFold 2 Beta Strand Database}
\begin{figure}
\centering
\includegraphics[width=\linewidth]{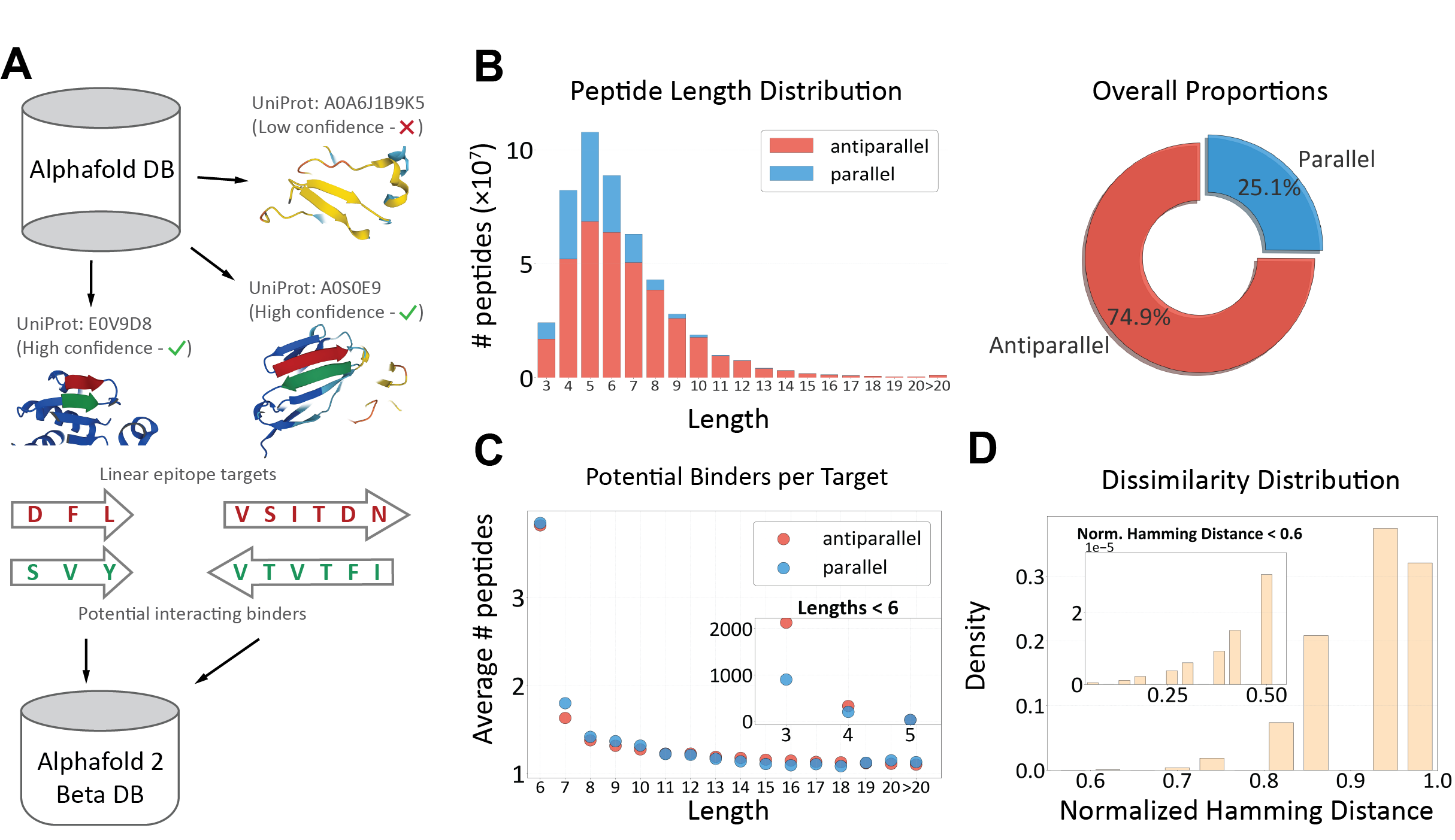}
\caption{\textbf{AlphaFold 2 Beta Strand Database.} (\textbf{A}) Illustration of the collection of high-confidence beta strand pairs from Alphafold Protein Structure Database. Example structures (viewed with Mol* viewer \protect\cite{sehnal2021mol}) show two high-confidence pairs (anti-parallel and parallel) and one pair that did not meet the high-confidence criteria.
(\textbf{B}) Peptide length distribution of beta strand pairs with overall proportion of anti-parallel (74.9\%) to parallel (25.1\%) pairs.
(\textbf{C}) Average number of potential binders available for each distinct target sequence. For clarity of visualization, lengths < 6 are plotted as an inset. (\textbf{D}) Pairwise dissimilarity distribution for anti-parallel length 8 data (a subset of 1,933,932 pairs) using the normalized Hamming distance. In \textbf{B}-\textbf{C}, pairs with lengths longer than 20 are grouped together for clarity.
}
\label{fig:Figure_Database}
\end{figure}

We constructed a large-scale database of sequence pairs sampling diverse beta-strand interaction motifs, building upon the release of Alphafold DB \cite{varadi2022alphafold} (Methods). We call this database AlphaFold 2 Beta Strand Database, which stores 488 million distinct beta strand pairs (Fig.~\ref{fig:Figure_Database}A). It shows a higher occurrence of antiparallel pairs, three times as many as parallel pairs (Fig.~\ref{fig:Figure_Database}B), which are potentially relevant to peptide design strategies due to stronger inter-strand stability and geometric planarity. The database contains data of significant diversity, with 97.7\% pairs of target-binder sequences showing less than 20\% similarity, as calculated by the normalized Hamming distance (Figs.~\ref{fig:Figure_Database}D, \ref{fig:Figure_Database_Supplementary}, Supp.~Methods~\ref{appendix:hamming}). Constructing large and diverse dataset is essential for training a deep language model, as machine translation performance was found to improve consistently with increasing dataset size \cite{gordon2021data}. This database is potentially useful for studying beta strand interactions in general and setting benchmarks for new prediction methods.

Despite the abundance and diversity of motifs in this database, especially at shorter lengths (Fig.~\ref{fig:Figure_Database}B), directly searching for matched epitope sequences and using the interaction partner as a binder candidate is often not a viable strategy. For a typical length 8 target, fewer than an average of 1.5 potential binders are available per target (Fig.~\ref{fig:Figure_Database}C). This highlights the need for extrapolating a general probability distribution over peptide sequence binding pairs through machine learning to deliver a broadly applicable design strategy and retrieve molecular patterns. 

\subsection{Deep learning model -- TransformerBeta}
To develop our machine learning design strategy, we formulated the peptide binder prediction problem as a machine translation task of beta strand interaction pairings, where one strand mimics a linear epitope target, while the other acts as a potential interacting binder, and we performed it through the Transformer architecture \cite{Vaswani2017}, which is the foundation for the majority of current language models. This choice builds upon the success of recent generative approaches for protein domain sequences \cite{meynard2023generating}, protein-specific drug molecules \cite{grechishnikova2021transformer} and signal peptide generation \cite{wu2020signal} based on casting the problem as a machine translation task and on the Transformer architecture.

\begin{figure}
\centering
\begin{subfigure}{0.9\textwidth}
  \centering
  \includegraphics[width=\linewidth]{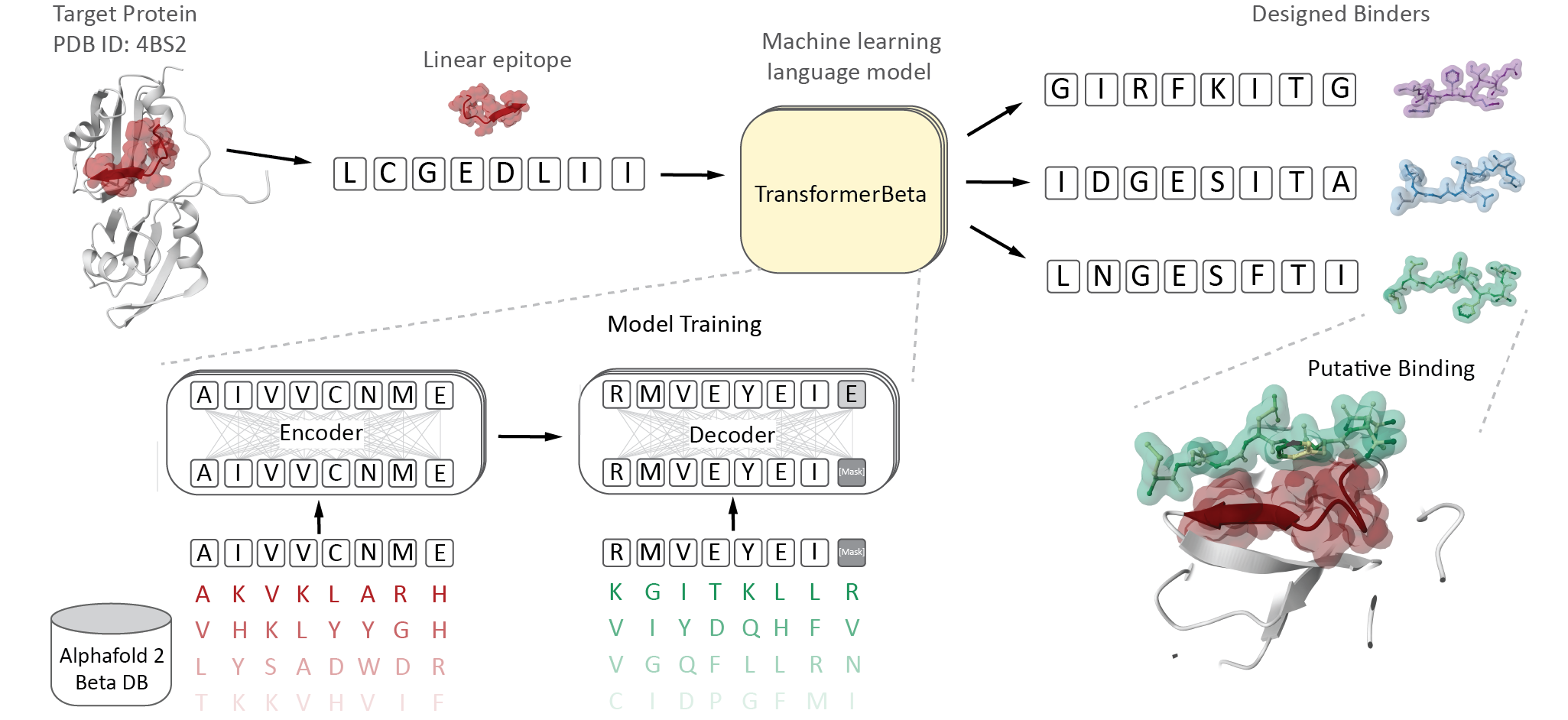}
\end{subfigure}
\caption{\textbf{Strategy for designing binders to target a linear epitope using TransformerBeta.} The target protein is shown in silver, with the epitope of interest highlighted in red. TransformerBeta takes as input the target sequence (from N to C terminus) and generates diverse target-specific linear peptides that are putative binders. The putative bound structure is a simulated docking pose using HPEPDOCK \protect\cite{zhou2018hpepdock} and all protein structures are viewed using Mol* viewer \protect\cite{sehnal2021mol}.}
\label{fig:Figure_strategy}
\end{figure}

For computational feasibility, we implemented our strategy on a curated dataset of 2.1 million length 8 antiparallel beta strand pairs from the AlphaFold 2 Beta Strand Database (Methods). Length 8 is the typical epitope length, and the designed peptide of this length is suitable for grafting onto a majority of antibody scaffolds by replacing the amino acids of CDR3 regions. However, our implementation is not constrained to peptides of length 8, allowing alignment-free training on varied input lengths. The Transformer model learns a probability distribution formulated autoregressively, which means that the probability of an amino acid at position $i$ ($y_i$, with $i=1,...,m$) depends on the previous amino acids ($y_{<i}$) and the input target sequence ($X$) (Methods, Supp.~Methods~\ref{appendix:model_architecture}, Fig.~\ref{fig:Figure_Database_Supplementary}). The probability of a target $X$-specific binder ($Y$) is the product of the likelihood of the individual amino acids $y_i$: 
\begin{equation}
\label{autoregressive_result}
P_{\theta}(Y|X)=P_{\theta}(y_1, ..., y_m|X)=\prod_{i=1}^{m} P_{\theta}(y_i|X, y_{<i})
\end{equation}
We performed training on 90\% of data by maximum likelihood, searched for optimal hyperparameters for a high-quality model on 5\% and evaluated its performance on the remaining 5\%. We then retrained the model with best hyperparameter settings on 100\% data for optimal performance (Supp.~Methods~\ref{appendix:model_selection}). As a result, we have TransformerBeta, a 6-layer encoder-decoder model with 44 million parameters (Fig.~\ref{fig:Figure_strategy}). Once evaluated on new peptides for a given target, TransformerBeta predicts a probability score, $P_{\theta}(Y|X)$, which represents the likelihood of the binder adopting a natural beta-strand conformation with the target, thus reflecting the peptide's quality as a target-specific binder. The designed binder could be used as a peptide candidate to carry on to additional computational and experimental tests in peptide and antibody design.

\subsection{TransformerBeta accurately predicts target-specific binders}
We tested the TransformerBeta's ability to accurately recover known binders for given target sequences. We assigned the TransformerBeta's probability score to target-binder pairs from the test set and from two negative-control datasets. In one of them, we coupled each target to random binders by concatenating amino acids uniformly at random (random set) to test the model's ability to capture the typical amino acid usage of beta-strand binders. In the second one, we shuffled the target-binder pairings, obtaining a randomly shuffled set that maintains all statistical properties but loses its target-specific nature (shuffled set), to test whether the model's predictions preserve information on specificity (Methods). The scores' distribution (Fig.~\ref{fig:probability_dist_analysis}A) shows the model's ability to discriminate the test data of genuine binders from both the shuffled and random data sets, assigning to the former a higher probability of target specificity. We quantified the model's performance using the Receiver Operating Characteristic (ROC) curve, the Precision-Recall (PR) curve, and the corresponding areas under the curve (ROC-AUC and PR-AUC respectively). The model achieved a ROC-AUC of 0.98 and PR-AUC of 0.98 for random peptides and ROC-AUC of 0.95 and PR-AUC of 0.96 for shuffled peptides, as shown in Fig.~\ref{fig:probability_dist_analysis}B-C, confirming that the model accurately distinguishes binders from non-binders. 

We further tested if TransformerBeta can well predict binders for \emph{unseen} target sequences. When constructing the test set, we specifically ensured that the test set contained no target-binder pair identical to those in the training set. Fig.~\ref{fig:probability_dist_analysis}D shows the performance on this test set, stratified by the closest Hamming distance of the test target relative to the training set target sequences: it stays high for unseen targets with distance 1 from training targets, with a decrease beyond this distance that however kept ROC-AUC>0.8 for distance 2 and ROC-AUC>0.65 for distances 3-4. The performance showed a similar pattern for distances between training and test binders (Fig.~\ref{fig:probability_dist_analysis}E). Generalization performance is expected to become increasingly difficult with higher distances between training and test set, as is well documented and typically controlled for in applications of ML to protein data \cite{tubiana2022scannet, hoie2024discotope, zhou2024antigen, kong2022conditional, luo2022antigen}. The prediction of immune receptors specific binding for unseen linear peptide targets is an extremely challenging and far from being solved \cite{jensen2024enhancing, meysman2023benchmarking, akbar2021compact}. In this context, for instance, it is already useful to achieve good predictive power for targets harbouring one mutation compared to the available data (like in the case of cancer neoantigens or viral single-point mutant epitopes); similarly, on the paratope side, 2-3 mutations can be sufficient to determine target specificity \cite{mayer2023measures, jensen2024enhancing, liu2022research}. For a fixed target, we also observed a trade-off performance-wise in terms of the heterogeneity of the corresponding binders in the training data (Fig.~\ref{fig:model_performance_supp}J-L), as it has been assessed in other contexts of statistical modelling of protein sequences \cite{posani2022infer, malinverni2023data}: target-specific training binders should be heterogeneous enough to sample their potential diversity, but not too divergent to preserve the necessary functionally-relevant statistical information. Such a result could help optimize training dataset construction from our AlphaFold 2 Beta Strand Database for specific targets. Finally, the model prioritizes binding pairs that appear frequently in beta strands and that have the potential to be promiscuous binders (Supp.~Methods~\ref{appendix: additional_assessment}, Fig.~\ref{fig:model_performance_supp}A-C, G-I). 

\subsection{TransformerBeta generates peptides similar to natural peptides}

For the peptide design task, it is key that the model generated sequences exhibit biophysical and functional properties akin to natural ones, ensuring \emph{e.g.}~high binding affinity and stability. We generated peptides with TransformerBeta (generated set) and compared them to a set of natural binders (test data binders of Section~\ref{sec:dataset_preparation}, referred to as the natural set) and a set of randomly concatenated amino acids (random set). First, to ensure that our generated peptides were not simply replicates of natural ones, we measured the Hamming distances between them for the same targets. We found an average Hamming distance of 4.6 with length 8 sequences, suggesting that the generated sequences were mostly novel. 

\begin{figure}
\centering
\includegraphics[width=0.9\textwidth]{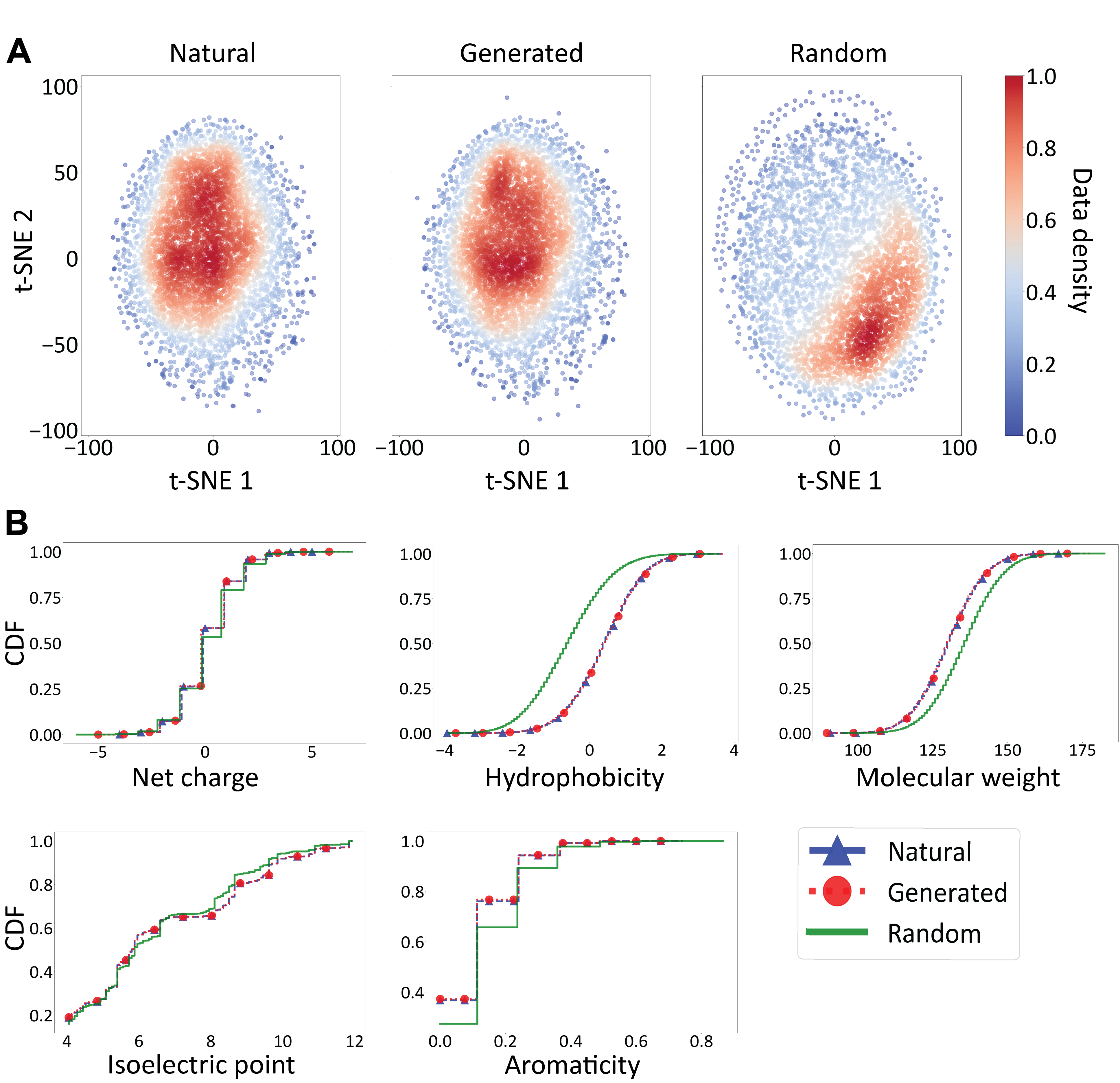}
\caption{\textbf{Properties of generated data.} 
(\textbf{A}) t-SNE projected distributions of 5,000 randomly sampled binders from natural set, the model generated set and random set. (\textbf{B}) Cumulative Distribution Function (CDF) of various physicochemical properties (Net charge, Hydrophobicity, Molecular weight, Isoelectric point, Aromaticity) for the same natural, model generated, and random binders as in (\textbf{A}).}
\label{fig:biophysical_properties}
\end{figure}

We projected a set of natural, generated and random peptides on a two-dimensional space using t-SNE \cite{van2008visualizing}(Fig.~\ref{fig:biophysical_properties}A). We observed a clear similarity in overall distribution between generated and natural sequences, but not with random ones, indicating that the model captures the organization of natural sequences in sequence space. Furthermore we found that generated sequences accurately reproduce the amino acid frequencies and correlations of natural sequences (Supp.~Methods~\ref{appendix:statistics_validation}, Fig.~\ref{fig:freq_check}), which is a fundamental test of the model's generative capacity \cite{morcos2011, russ2020, mcgee2021generative, sgarbossa2023generative}.

Finally, we focused on the physicochemical properties known to influence peptide stability and molecular interactions \cite{lee2023strategies, bak2015physicochemical}. By comparing the cumulative distribution functions (CDF) of five physicochemical properties (Net charge, Hydrophobicity, Molecular weight, Isoelectric point and Aromaticity) across different data groups, we observed that generated and natural peptides have a substantial overlap in terms of these properties, while being clearly distinct from random sequences (Fig.~\ref{fig:biophysical_properties}B). The differences in these properties' distribution between natural and generated peptides are not statistically significant in four out of five cases (p-value $>0.05$, Kolmogorov's D-statistic tests, Supp.~Methods \ref{appendix:physicochemical_validation}), with a slight deviation only in aromaticity. We specifically noted that TransformerBeta emulates the hydrophobicity distribution of natural peptides, which is substantially higher than that of random sequences (Fig.~\ref{fig:biophysical_properties}B). This is expected since beta sheets often form the hydrophobic core of proteins, but it implies the model would design binders with hydrophobic tendencies potentially compromising their solubility, which should be additionally screened by tailored computational methods \cite{magnan2009, sormanni2015, khurana2018}.  
\subsection{Model's biological interpretability}

\begin{figure}
\centering
\includegraphics[width=0.99\linewidth]{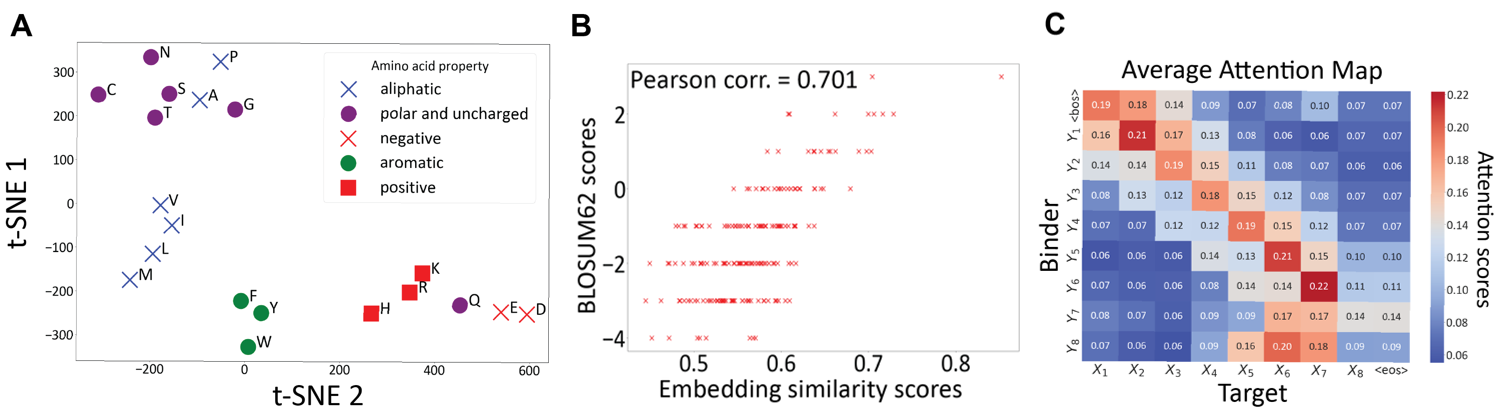}
\caption{\textbf{Interpretability of TransformerBeta.} 
(\textbf{A}) Input embedding shared across encoder and decoder. 
(\textbf{B}) Scatter plot comparing embedding cosine similarity scores and BLOSUM62 substitution scores (Supp.~Methods~\ref{appendix: embedding}). 
(\textbf{C}) Average cross-attention map. $X_i$ and $Y_i$ represents the $i^{th}$ amino acid of target and binder respectively. <bos> and <eos> are two special tokens.} 
\label{fig:embedding_analysis}
\end{figure}

Language models have been shown to learn representations that recapitulate interpretable biological information and capture key protein features, including structure, function, interactions and evolution \cite{rives2021biological, elnaggar2022prottrans, vig2020bertology,leem2022deciphering}. To check whether the learnt representations of TransformerBeta are biologically sensible, we first studied its input embedding layer, a dictionary of 20 learnable amino acid vectors shared across the encoder and decoder (Fig.~\ref{fig:embedding_analysis}A). We projected the high-dimensional embedding vectors onto a two-dimensional space using t-SNE to visualize the distribution of amino acids, annotating them by the main physico-chemical property, including charge, polarity, hydrophobicity and aromaticity (Fig.~\ref{fig:embedding_analysis}A). The trained embedding displayed clear clustering patterns corresponding to such groups, demonstrating the model has captured biologically meaningful representations in the embedding space, similar to other language models \cite{rives2021biological, elnaggar2022prottrans}. We compared the TransformerBeta's embedding layer with the amino acid substitution propensities summarized by the BLOSUM62 matrix \cite{henikoff1992amino} (reflecting physico-chemical similarities), and with the learnt embeddings of established pretrained protein language models \cite{elnaggar2022prottrans, rao2019evaluating}, finding a relatively high degree of correlation with all of them (Figs.~\ref{fig:embedding_analysis}B, \ref{fig:embedding_analysis_benchmark}, Table~\ref{tab:protein_models_comparison}, Supp.~Methods \ref{appendix: embedding}).  

Finally, we assessed if our model has learned features specific to our beta strand data, beyond general amino acid properties. We extracted all cross-attention maps for a set of 1,000 target-binder pairs randomly sampled from the training data (Supp.~Methods~\ref{appendix:model_architecture}). These attention values reflect the relevance of each amino acid in the target for the prediction, and hence the design, of each amino acid in the binder. In average, attention values are mostly concentrated along the diagonal (Fig.~\ref{fig:embedding_analysis}C): when predicting the next amino acid of the binder, the model leverages maximally the information from the facing residue along the target, in line with the fact that hydrogen bonds between facing residues are the key interactions in beta strands \cite{wouters1995analysis}.

\begin{figure}
\centering
\includegraphics[width=0.80\linewidth]{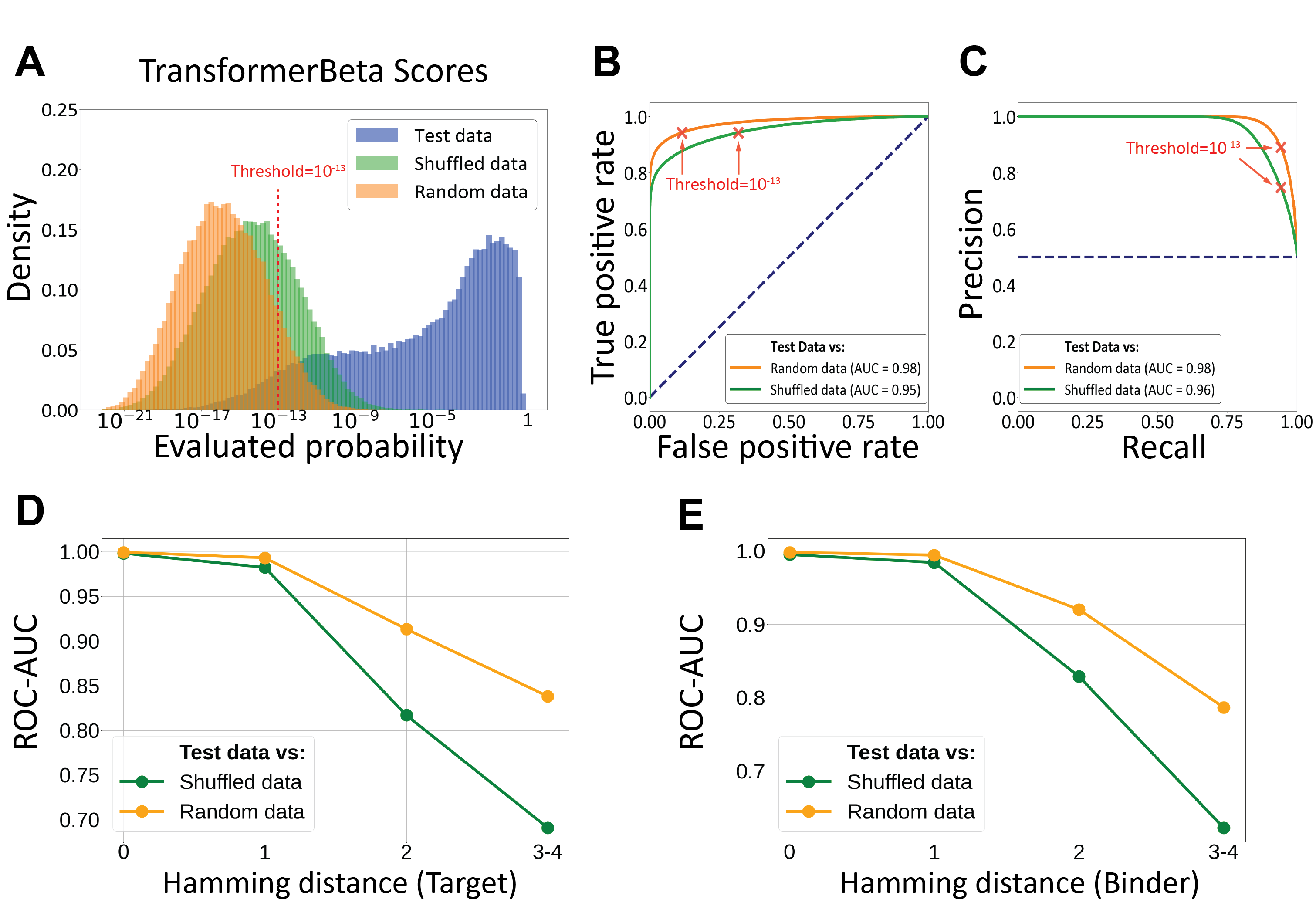}
\caption{\textbf{Model's prediction accuracy.} (\textbf{A}) Distribution of the TransformerBeta predicted probability assigned to binders in the test, shuffled, and random sets, each containing 107,441 sequence pairs. (\textbf{B}) Receiver Operating Characteristic (ROC) curve and corresponding Area Under the Curve (ROC-AUC). (\textbf{C}) Precision-Recall (PR) curve and corresponding Area Under the Curve (PR-AUC). The dashed line gives the performance of a random classifier (ROC-AUC$=$PR-AUC$=$0.50). ROC-AUC as a function of the closest Hamming distance between test and training targets (\textbf{D}) and between test and training binders (\textbf{E}). Hamming distances of 4, having fewer than 5 data points, are grouped with distance 3.}

\label{fig:probability_dist_analysis}
\end{figure}

\section{Conclusion}
In this paper, we built upon recent progress in machine learning architectures for language translation and text generation to train a generative model of linear peptide binders (TransformerBeta) on a dictionary of beta strand pairs. To obtain a more exhaustive sampling of such interaction motifs for the model's training, we leveraged the predicted structures made available by the Alphafold Protein Structure Database \cite{varadi2022alphafold}. We have shown that TransformerBeta recovers with high accuracy complementary beta strands, and that by sampling from the learnt distribution one can generate novel candidate peptides with beta strand characteristics that resemble natural ones statistically and physico-chemically. The generative power of TransformerBeta could be exploited to design high-affinity beta-strand-like binders specific to pre-selected linear epitopes. We found that TransformerBeta learns representations recapitulating beta strand-specific binding modes (like the presence of hydrophobic face-to-face bonds) and general amino acid properties. 

We provided proof-of-concept of our TransformerBeta design strategy for peptide pairs of fixed length, nevertheless future efforts could improve its ability to generalize to varying lengths by exploiting more data from our AlphaFold 2 Beta Strand Database, albeit at the cost of increased computational power. Systematic comparisons with existing peptide design methods and computational pipelines \cite{sormanni2015rational, bryant2022evobind, malinverni2023data} will be needed in connection to specific design applications. Our beta strand database could also be further expanded with new predicted structures becoming available, \emph{e.g.}, following the release of AlphaFold3 \cite{abramson2024accurate}.

Converting structural motifs into a library of interacting sequence fragments gave the advantage of building an approach that is sequence-based, hence computationally cheaper, making it useful to score or generate large numbers of putative binders in a short time. TransformerBeta could be used to select a few high-quality starting sequences for in vitro optimization of ligands, or to generate epitope-specific peptide libraries for experimental screening to accelerate drug discovery. It could be flexibly coupled with functional motif scaffolding methods for the design of a binding protein, \emph{e.g.}~in antibody engineering, where designed peptides are grafted onto the antigen-binding region of an antibody scaffold. Finally, it could serve as baseline model of generic beta-sheet-type interactions to train via transfer learning models of specific types of interactions with linear targets and ligands, for instance some immune epitope-paratope interactions \cite{weber2021, bennett2024atomically}.


\section{Methods}

\subsection{Creation of AlphaFold 2 Beta Strand Database} \label{method: database_construction}
We developed the AlphaFold 2 (AF2) Beta Strand Database to store high-confidence scored beta strand pairs predicted by Alphafold 2 \cite{jumper2021highly}. We downloaded all 214 million protein structures from the Alphafold Protein Structure Database (Alphafold DB) \cite{varadi2022alphafold} with data access dated 07/10/2022, available at \texttt{https://alphafold.ebi.ac.uk/}. We extracted the equal-length pairs of amino acid sequences that face each other in a beta-strand conformation with the following criteria: (i) Both sequence lengths are at least 3 residues; (ii) All residues have a pLDDT of at least 70; (iii) All residues that face each other have a PAE of less than or equal to 10. We obtained a total of 1,532,767,459 beta strand pairs from the Alphafold DB and reduced them to 488,153,783 unique beta strand pairs, each with a count value indicating its frequency in the original 1.5 billion dataset. The database is available at \texttt{https://huggingface.co/datasets/hz3519/AF2\_Beta\_Strand\_Database}.

\subsection{Dataset preparation for model training and evaluation}
\label{sec:dataset_preparation}

We retrieved 38,616,564 length 8 antiparallel beta strands from the AF2 Beta Strand Database. We removed data with counts less than 5, resulting in a final dataset with 2,148,813 pairs. The dataset was randomly split into training data, comprising 90\% of the data (1,933,932 pairs), validation data, with 5\% (107,440 pairs) and test data, with the remaining 5\% (107,441 pairs). 

For evaluation, we generated additional control datasets, which we refer to as `random' or `shuffled'. Random binders were generated by uniformly sampling from the 20 amino acids at each position for each test target. Shuffled binders were obtained by permuting the order of true binders for the test set. Samples of `generated' binders were generated by a probabilistic sampling strategy (Supp.~Methods~\ref{appendix:model_selection}) for each test target. We constructed the final random, shuffled and generated sets, each containing 107,441 pairs, by pairing the binders with corresponding test targets. 

\subsection{Generative model for peptide design -- TransformerBeta}

Our aim was to learn the conditional probability distribution $P(Y|X)$ of natural beta strand space, where $X$ is the target and $Y$ is the binder. We treated the amino acids as discrete tokens and represented $X$ and $Y$ as sequences of tokens $(x_1, \ldots, x_n)$ and $(y_1, \ldots, y_m)$, respectively. Given that AF2 Beta Strand Database contains target and binder pairs of equal length, the lengths of the target and binder sequences are equal ($n=m$). We trained a standard Transformer \cite{Vaswani2017} (Supp.~Methods~\ref{appendix:model_architecture}, \ref{appendix: hardware}), an autoregressive model that learns the probability $P_{\theta}(Y|X)$ given by Equation~\ref{autoregressive_result}, then used for the beta-strand-like binders generation task (Supp.~Methods~\ref{appendix:model_selection}). The parameters $\theta$ specifying the conditional probability distribution $P_{\theta}(Y|X)$ are learnt by maximizing the log-likelihood over the training data $\mathcal{D_T}$, i.e., by finding $\theta^{*}$ such that:
\begin{equation}\theta^{*} = \underset{\theta}{\operatorname{argmax}}\sum_{(X, Y) \in \mathcal{D_T}}\sum_{i=1}^{m}\log P_{\theta}(y_i|X, y_{<i})
\label{max_log_likelihood}
\end{equation}
The code is available at \texttt{https://github.com/HZ3519/TransformerBeta\_project}. 
\clearpage
\section*{Acknowledgments}
FAA thanks UK Research and Innovation (Future Leaders Fellowship MR/S033947/1 and MR/Y003616/1) for support.

\bibliography{neurips_2024}
\bibliographystyle{unsrt}

\clearpage
\section{Supplementary Methods} 

\subsection{Database dissimilarity measures} 
\label{appendix:hamming}

To estimate the dissimilarity in peptide composition of our dataset, we concatenated the sequence pairs and calculated the pairwise normalized Hamming distance between different concatenated sequences (each of length 16). The Hamming distance, denoted as $H(x, y)$, is defined as the number of positions at which the corresponding amino acids in sequences $x$ and $y$ are different. The normalized Hamming distance is given by the formula:

\begin{equation}
\text{Normalized Hamming distance} = \frac{H(x, y)}{16}
\end{equation}

\begin{figure}[htbp]
\centering
\includegraphics[width=1\textwidth]{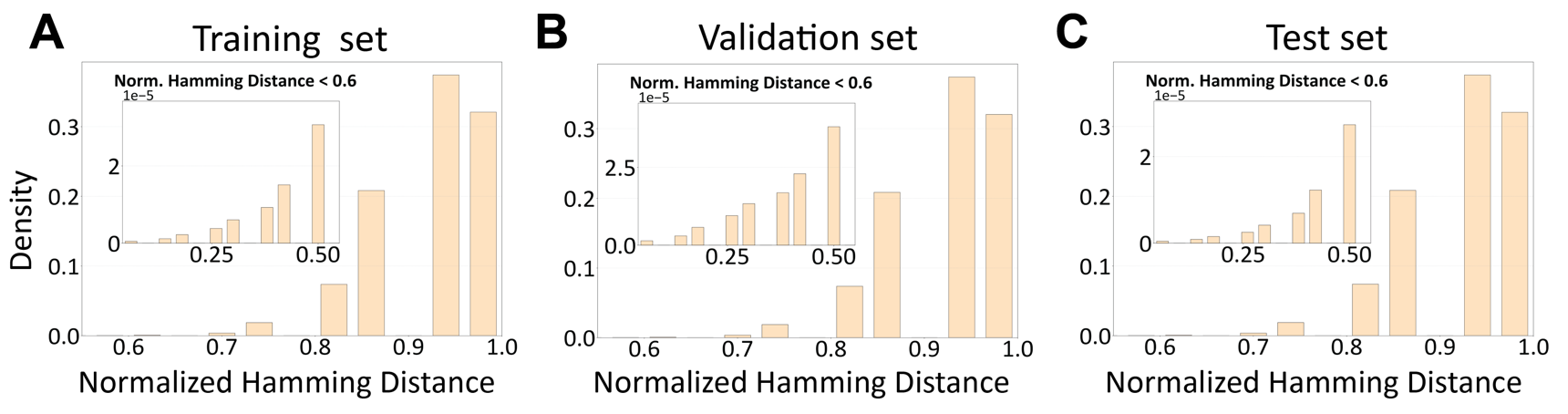}
\caption{\textbf{Pairwise dissimilarity distributions for concatenated sequences in training, validation and test sets.} Distribution of normalized Hamming distance within: (\textbf{A}) a randomly sampled portion of the training set (10\%, 193,393 sequences); (\textbf{B}) the validation set (107,440 sequences); (\textbf{C}) the test set (107,441 sequences). The score ranges between 0 and 1, with higher values indicating greater dissimilarity. For clarity of visualization, a zoom on the density for normalized Hamming distance < 0.6 is provided in the insets. Panel \textbf{D} of Fig.~\ref{fig:Figure_Database} is the training set distribution in panel \textbf{A} here.}
\label{fig:Figure_Database_Supplementary}
\end{figure}

The sequence pairs within all datasets (training, validation, and test) were sufficiently diverse for model training and evaluation as the majority of sequence pairs exhibited less than 20\% similarity (Figs.~\ref{fig:Figure_Database}, \ref{fig:Figure_Database_Supplementary}).

\subsection{Model architecture}\label{appendix:model_architecture}

\begin{figure}[htbp]
\centering
\includegraphics[width=1\textwidth]{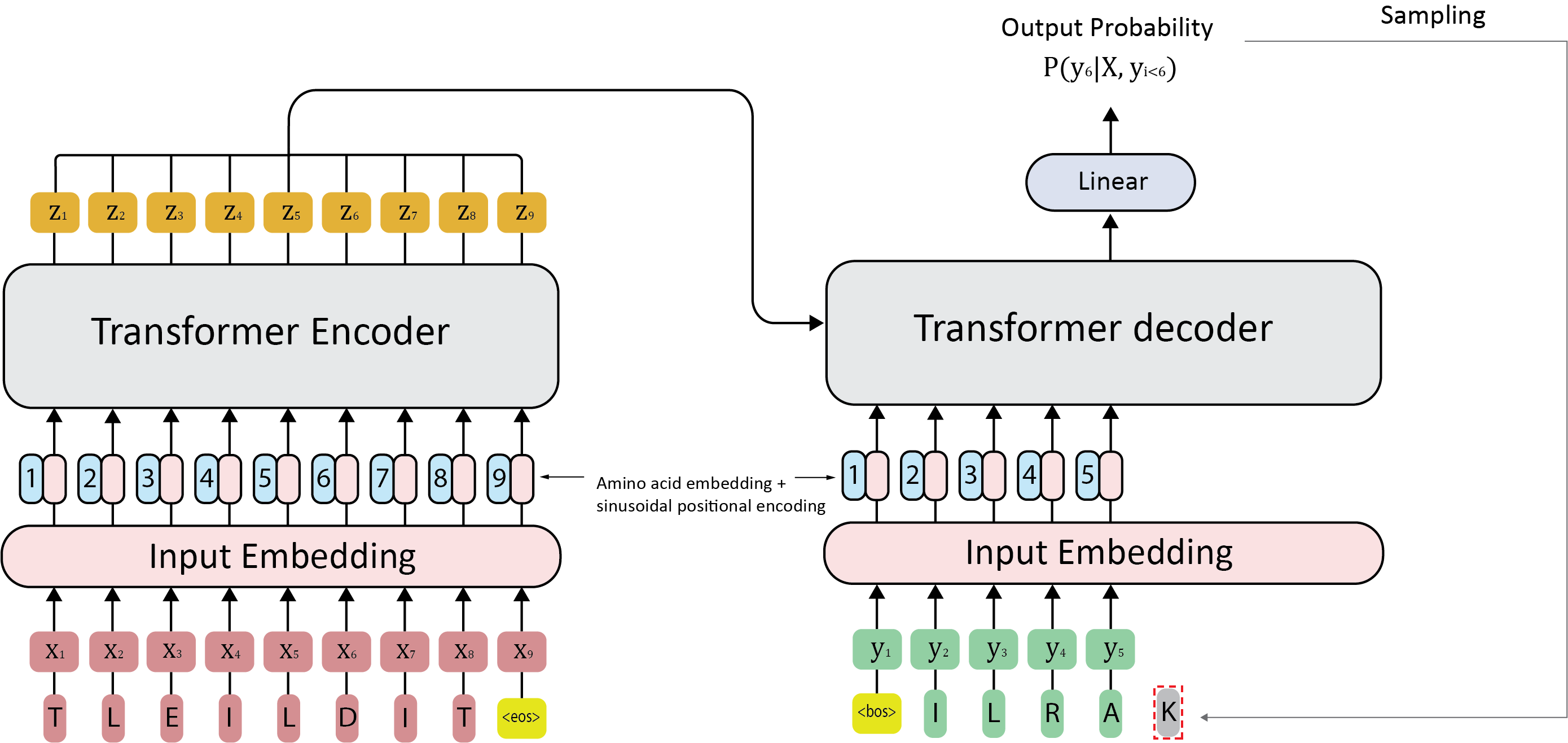}
\caption{\textbf{TransformerBeta architecture and sampling strategy.} The figure depicts the generation of a binder "ILRAKVIL" for the target sequence "TLEILDIT" at the step of decoding symbol in position 6 along the sequence. Red tokens represent the target sequence, green tokens represent the decoded sequence for previous steps and gray token represents the next decoded token in this example. <eos> and <bos> are 2 special learnable tokens such that the decoding process starts with <bos> token and continues until the binder with a length equal to the target sequence is generated. 
}
\label{fig:decoding_animation}
\end{figure}

To learn the autoregressive model of Equation~\ref{autoregressive_result}, we employed the encoder-decoder neural network architecture known as Transformer, as illustrated in Fig.~\ref{fig:decoding_animation}. We adopted the same architecture in the original paper \cite{Vaswani2017}, where both the encoder and the decoder are composed of a stack of $N_{L}$ identical layers. The Transformer architecture, its training and performance have been revised in detail in several studies \cite{phuong2022formal, LIN2022111, Tay2022}. For the sake of a brief illustration, we describe the structure of a typical Transformer encoder layer containing two key components: a multi-head self-attention module followed by a position-wise fully connected feed-forward neural network. We omit the details of the decoder, as it closely resembles the encoder. 


Before entering the encoder, the target sequence of amino acids $X = (x_1,...,x_n)$ is first preprocessed by an input embedding transformation and positional encoding. Each amino acid is independently transformed by a learnable input embedding layer into a vector of dimension $d_{model}$. The embedded vectors are added by the sinusoidal positional encoding as in \cite{Vaswani2017}, which produces a matrix $E$ of dimension $n \times d_{model}$. The transformation of input embedding layer is learnt during the training process and gives a representation of each amino acid as vector of continuous values. We quantify the correlation between two amino acids using embedding similarity, calculated as the cosine similarity between their embedding representations $\vec{a}$ and $\vec{b}$, given by the following formula:

\begin{equation}
\text{Embedding similarity}(\Vec{a}, \vec{b}) = \text{Cosine  similarity}(\vec{a}, \vec{b}) = \frac{\vec{a} \cdot \vec{b}}{||\vec{a}||_2 \cdot ||\vec{b}||_2}
\label{eq:embedding_similarity}
\end{equation}

The multi-head attention module then projects in parallel the embedded input $E$ into $h$ sets (or heads) of query ($Q$), key ($K$), and value ($V$) representations through learnable linear transformations. For each head $i=1, \dots, h$, the matrices $Q$, $K$, and $V$ have dimensions $d_{model}\times d_q$, $d_{model}\times d_k$, and $d_{model}\times d_v$, respectively; here we chose $d_q=d_k = d_v = d_{model}/h$ as in \cite{Vaswani2017}. The attention function for each of the $h$ heads is computed as follows:
\begin{equation}
\operatorname{Attention}^{(i)}(Q^{(i)}, K^{(i)}, V^{(i)}) = \operatorname{softmax}\left(\frac{Q^{(i)}K^{(i)T}}{\sqrt{d_{model}/h}}\right)V^{(i)}, \qquad i=1, \dots, h
\label{eq:attention}
\end{equation}
The $n \times n$ matrix $\operatorname{softmax}\left(\frac{Q^{(i)}K^{(i)T}}{\sqrt{d_{model}/h}}\right)$ provides, for a given input sequence, the weights by which each token attends to the other tokens along the sequence, quantifying its relevance to their representation and prediction and hence reflecting the degree of statistical interdependence among tokens. The $h$ attention outputs \ref{eq:attention} (each of dimension $n \times d_{model}/h$) are concatenated to produce a multi-head attention output with dimension $n \times d_{model}$. 

Next the position-wise feed-forward neural network applies, independently to each position of this multi-head attention output, two linear transformations with a ReLU activation in the middle. The first transformation projects the output to dimension $n \times d_{ff}$, while the second transformation projects it back to the original dimension $n \times d_{model}$, producing the final output of a Transformer encoder layer - a high-level protein feature representation vector $Z = (z_1, \ldots, z_n)$. The encoded representations $Z$ then enter the cross attention module in the decoder, which operates similarly to Equation~\ref{eq:attention}. In the cross attention module, the value ($V$) projections are derived from $Z$, while query ($Q$) and key ($K$) projections are derived from binder sequences. 

During training, the model outputs a predicted probability distribution $P_{\theta}(y_i|X, y_{<i})$ over 20 amino acids at each decoding position $i$. This predicted probability is compared to the true amino acid at the same decoding position, which is represented by a categorical distribution $q(y_i|X, y_{<i})$ over the 20 amino acids, with probability 1 at the correct amino acid and 0 at others. The Transformer learns the parameter set $\theta^{*}$ that minimizes the categorical cross-entropy loss across all decoding positions in the training dataset $\mathcal{D_T}$:

\begin{equation}
\theta^{*} = \underset{\theta}{\operatorname{argmin}}\left(-\sum_{(X, Y) \in \mathcal{D_T}}\sum_{i=1}^{m}\sum_{k\in \mathcal{S}} q(y_i=k|X, y_{<i}) \log P_{\theta}(y_i=k|X, y_{<i})\right)
\label{min_cross_entropy}
\end{equation}

where $k$ represents a specific amino acid within the 20-amino-acid set $\mathcal{S}$. The minimization of the cross-entropy loss aims to align the model's predicted amino acid frequencies (via $P_{\theta}$) with the empirical ones (via $q$). This process is equivalent to maximizing the log-likelihood (Equation~\ref{max_log_likelihood}). As in \cite{Vaswani2017}, we used two regularization techniques to prevent overfitting during training: applying a dropout with probability $P_{drop}$ to the output of each sub-layer and label smoothing $\epsilon_{ls}$ to the cross-entropy loss \cite{szegedy2016rethinking}.

\subsection{Model selection}\label{appendix:model_selection}

We evaluated the performance of Transformer models with various hyperparameter configurations by changing the number of layers ($N_{L}$), the dimension of embedding ($d_{model}$), the dimension of the feed-forward layer ($d_{ff}$), the number of heads for the multi-head attention module ($h$), the dropout probability ($P_{drop}$), label smoothing ($\epsilon_{ls}$), and the number of training steps (Table~\ref{tab:model_parameters}). We used Adam optimizer \cite{kingma2014adam} with $\beta_{1}=0.9$, $\beta_{2}=0.98$ and $\epsilon = 10^{-9}$ (the default choices by \cite{Vaswani2017}). A cosine warmup schedule was utilized for the Adam optimizer, incorporating a linear increase warmup phase over 10,000 steps, for training stabilization \cite{liu2019variance, popel2018training}. We implemented embedding sharing for encoding input embedding and decoder input embedding, which was shown to achieve similar (or better) performance while effectively reducing the model's parameters \cite{lan2019albert, reid2021subformer}. We maintained a consistent batch size of 4096 and a learning rate of 0.004 across the hyper-parametric search. 

The primary metric used for model selection was the average log-likelihood of the validation dataset $\mathcal{D_V}$, defined as: 

\begin{equation}
\mathcal{L_V} = \frac{1}{|\mathcal{D_V}|} \sum_{(X, Y) \in \mathcal{D_V}}\log P_{\theta^{*}}(Y|X)
\label{validation_log_likelihood}
\end{equation}

which measures the model's capability to assign high probabilities to unseen data. We supplemented this primary evaluation metric with additional metrics to ensure the quality of the selected model. To ensure the selected model would capture the statistical properties of beta strands, we calculated the single-site frequency and 2-point connected correlations for both generated and validation binders, monitoring the Mean Absolute Errors ($MAE_{1}$, $MAE_{2}$) on these quantities for each model (Supp.~Methods~\ref{appendix:statistics_validation}, Fig.~\ref{fig:freq_check}). To confirm the selected model's ability to capture the physicochemical properties of beta strands, we looked at the distributions of various physicochemical properties (charge, hydrophobicity, molecular weight, isoelectric point and aromaticity) for both generated and validation binders, and calculated the Kolmogorov's D-statistic, representing the maximum absolute difference between two empirical distributions (Supp.~Methods~\ref{appendix:physicochemical_validation}). 

As a final model to generate results, we have selected Model M in Table \ref{tab:model_parameters}, as it demonstrated the best performance in terms of $\mathcal{L_V}$. Across two additional statistical metrics, this model proves to be the best performing in terms of $MAE_1$ and third best performance in terms of $MAE_2$. Furthermore, 4 out of 5 Kolmogorov-Smirnov property tests cannot be rejected at the 0.05 significance level, \emph{i.e.}, the distribution of the corresponding physicochemical properties across the generated sequences is not significantly distinguishable from the one across the sequences of the validation set. We used Model M trained on 90\% of the training data for subsequent validation analyses. For the generation of peptides, we retrained the parameter settings of Model M using the full dataset (2,148,813 pairs, Methods~\ref{sec:dataset_preparation}) for optimized performance.

To generate peptides for evaluation, we adopted a random sampling strategy that sequentially generating amino acids for the binder given a target. Amino acids are sampled from the learnt conditional distribution \eqref{autoregressive_result} until both the binder and target sequence attain equal length. 

\subsection{Additional assessment of model performance}\label{appendix: additional_assessment}

\begin{figure}[htbp]
\centering
\includegraphics[width=1\textwidth]{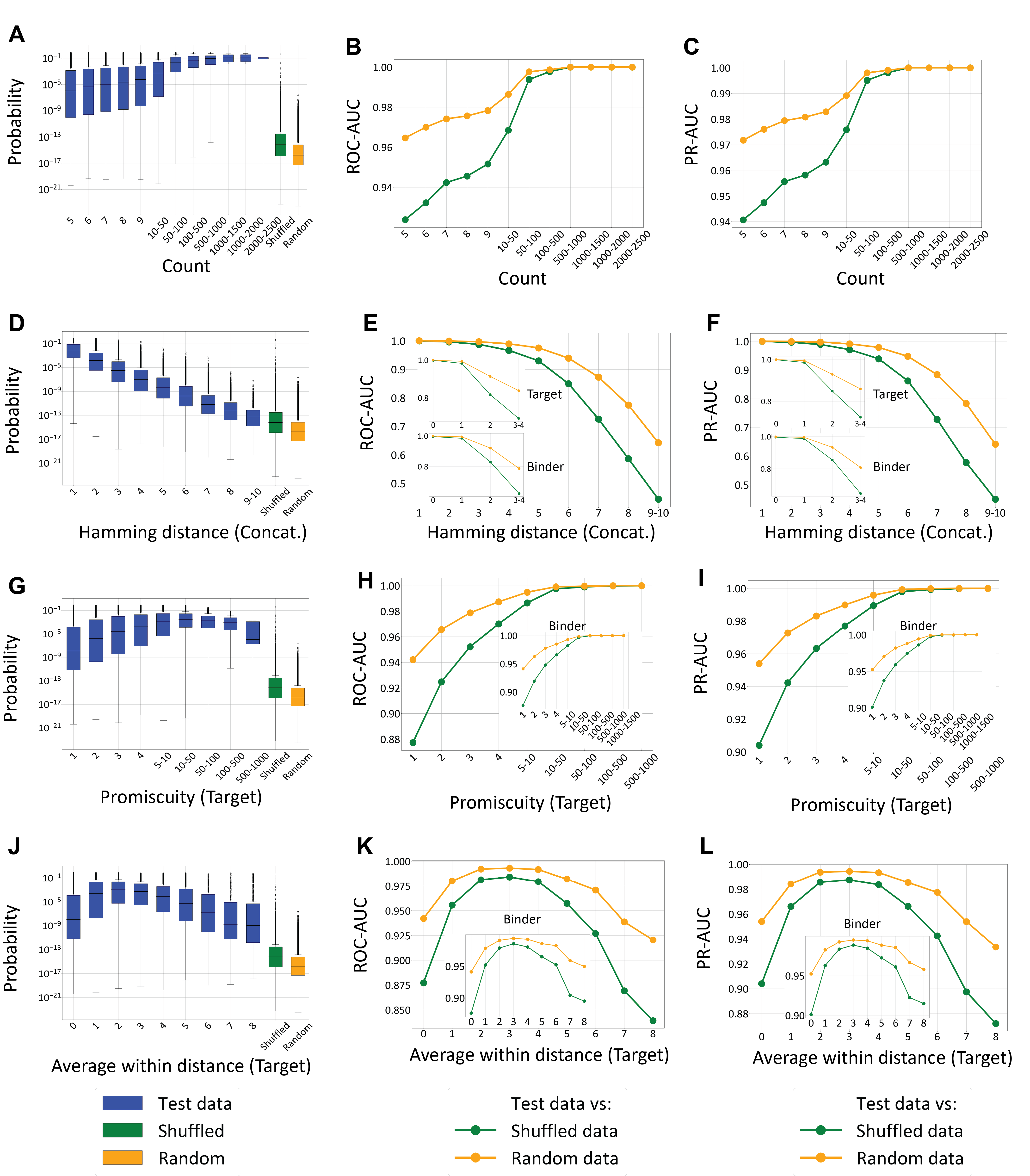}
\caption{\textbf{TransformerBeta's ability to discriminate test data from shuffled and random data with varying factors}. The average Area Under the Receiver Operating Characteristic Curve (ROC-AUC) is plotted as a function of: (\textbf{B}) occurrence count of each test data point in AF2 Beta Strand database; (\textbf{E}) minimum Hamming distance between concatenated sequences (targets and binders as insets) in the test data and closest training data; (\textbf{H}) promiscuity score of test targets (binders as inset) when searched in AF2 Beta Strand database; (\textbf{K}) average pairwise Hamming distance of binders for each test target (targets for each binder as inset) when searched in AF2 Beta Strand database. Similar Area Under the Precision-Recall Curves (PR-AUC) are monitored in (\textbf{C}), (\textbf{F}), (\textbf{I}) and (\textbf{L}). The probability distributions generating the AUC plots are shown in (\textbf{A}), (\textbf{D}), (\textbf{G}) and (\textbf{J}).}
\label{fig:model_performance_supp}
\end{figure}

In this section, we define the conditions informative about model performance assessed in Figs.~\ref{fig:probability_dist_analysis} and \ref{fig:model_performance_supp}.

\begin{enumerate}
\item "Hamming distance (Target)" represents the minimum Hamming distance target sequences in the test data and the closest training data. 
\item "Hamming distance (Binder)" represents the minimum Hamming distance binder in the test data and the closest training data. 
\item "Count" represents the frequency of occurrence of each test data point in the AF2 Beta strand database.
\item "Hamming distance (Concat.)" represents the minimum Hamming distance between concatenated sequences, by joining target sequences and binder sequences, in the test data and the closest training data. 
\item "Promiscuity (Target)" represents the number of binders for a test target in the AF2 Beta Strand database. 
\item "Promiscuity (Binder)" represents the number of targets for a test binders in the AF2 Beta Strand database. 
\item "Average within distance (Target)" represents the average pairwise Hamming distance of binders for a test target in AF2 Beta Strand database. 
\item "Average within distance (Binder)" represents the average pairwise Hamming distance of targets for a test binder in AF2 Beta Strand database. 
\end{enumerate}

We additionally monitored model performance against other metrics: `Count' represents the frequency of target-binder pairs; `Promiscuity' is given by the number of binders per target (and vice versa) and is a proxy for a peptide's versatility in binding; `Average within distance' quantifies the sequence diversity of binding partners to the same peptide. In conclusion, we found that TransformerBeta predictions have higher accuracy for higher count, higher promiscuity and medium average within distance, as shown in Fig.~\ref{fig:model_performance_supp}, essentially reflecting the level of sampling of target-binder pairs in the training set. Beyond sampling, a space of binding partners too narrow or too heterogeneous leads to slightly less accurate predictions (Figs.~\ref{fig:model_performance_supp}J-L).  

\subsection{Validation on statistical properties}
\label{appendix:statistics_validation}

The statistical covariation between sequence positions encodes key evolutionary protein information and is highly relevant in generating functional synthetic sequences, as experimentally demonstrated by previous research \cite{russ2020}. We evaluated three statistical properties, namely the single-site amino acid frequency $f_i(a)$, \emph{i.e.}~the frequency of the amino acid $a$ at position $i$ along the sequence in the sample under consideration, the 2-point connected correlation $C_{ij}(a, b)$, and the 3-point correlation $C_{ijk}(a, b, c)$. The definitions are as follows. The two-point connected correlation of amino acids $a$ and $b$ at distinct positions $i$ and $j$, respectively, is given by:
\begin{equation}
C_{ij}(a, b) = f_{ij}(a, b) - f_{i}(a)f_{j}(b)
\label{eq:two_point_correlation}
\end{equation}
where $f_{ij}(a, b)$ denotes the joint frequency of amino acids $a$ and $b$ occurring at positions $i$ and $j$ (where \(i \neq j\)), respectively. $C_{ij}(a, b)$ measures the degree to which the observed frequency of $a$ and $b$ appearing together deviates from what would be expected if their occurrences were independent.

The three-point correlation of amino acids $a$, $b$, and $c$ at distinct positions $i$, $j$, and $k$, respectively, is given by: 
\begin{equation}
C_{ijk}(a, b, c) = f_{ijk}(a, b, c) - f_i(a)f_{jk}(b, c) - f_j(b)f_{ik}(a, c) - f_k(c)f_{ij}(a, b) + 2f_i(a)f_j(b)f_k(c)
\label{eq:three_point_correlation}
\end{equation}
where $f_{ijk}(a, b, c)$ denotes the joint frequency of amino acids $a$, $b$, and $c$ occurring at positions $i$, $j$, and $k$ (where \(i \neq j\), \(i \neq k\), and \(j \neq k\)), respectively. It measures the degree to which the observed frequency of amino acids $a$, $b$, and $c$ appearing together at positions $i$, $j$, and $k$ deviates from what is expected if their occurrences were independent or only pairwise dependent. The three-point correlation provides a higher-order statistical description of the beta strand sequence space, capturing the intricate interdependence among the amino acid residues. 

\begin{figure}[htbp]
\centering
\includegraphics[width=1\textwidth]{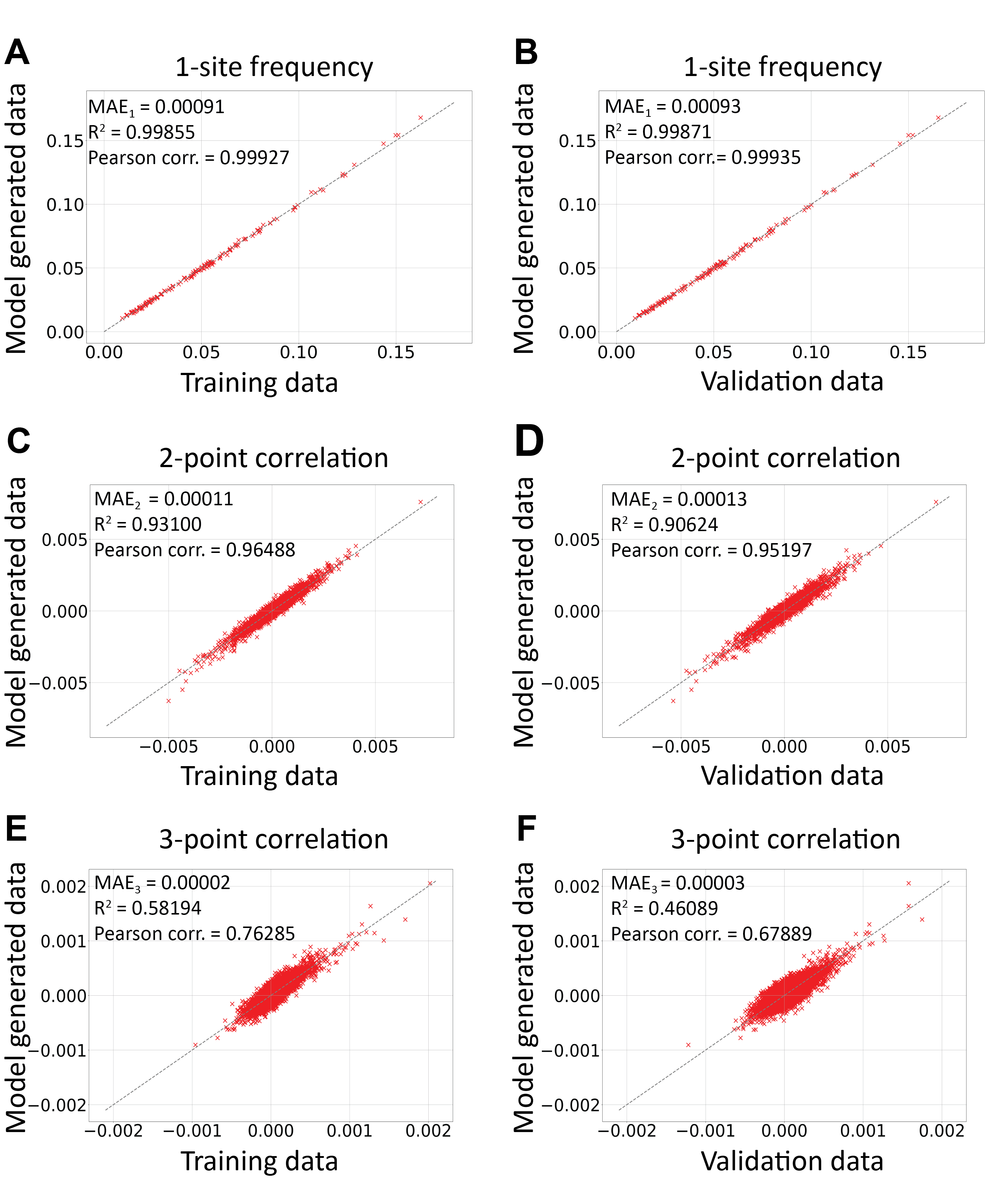}
\caption{\textbf{Model's validation on statistical properties.} Comparison of 1-site frequency (\textbf{A-B}), 2-point connected correlations (\textbf{C-D}) and 3-point connected correlations (\textbf{E-F}) between model generated binders (107,440 sequences) and: training binders (1,933,932 sequences, \textbf{A}, \textbf{C}, \textbf{E}); validation binders (107,440 sequences, \textbf{B}, \textbf{D}, \textbf{F}). The model used to generate new samples is Model M (Table \ref{tab:model_parameters}).}
\label{fig:freq_check}
\end{figure}

The Mean Absolute Errors (MAEs) between the original data statistics ($f_i(a)$, $C_{ij}(a, b)$, and $C_{ijk}(a, b, c)$) and the model predicted ones ($f_i^{\prime}(a)$, $C_{ij}^{\prime}(a, b)$, and $C_{ijk}^{\prime}(a, b, c)$) are computed as follows:
\begin{equation}
MAE_{1} = \frac{1}{m|\mathcal{S}|}\sum_{i=1}^{m}\sum_{a \in \mathcal{S}}|f_i(a) - f_i^{\prime}(a)|
\label{eq:mae_1}
\end{equation}
\begin{equation}
MAE_{2} = \frac{2}{m(m-1)|\mathcal{S}|^2}\sum_{\substack{i=1, j=1\\i\neq j}}^{m}\sum_{a, b \in \mathcal{S}}|C_{ij}(a, b) - C_{ij}^{\prime}(a, b)|
\label{eq:mae_2}
\end{equation}
\begin{equation}
MAE_{3} = \frac{6}{m(m-1)(m-2)|\mathcal{S}|^3}\sum_{\substack{i=1, j=1, k=1\\i \neq j,i \neq k,j \neq k}}^{m}\sum_{a, b, c \in \mathcal{S}}|C_{ijk}(a, b, c) - C_{ijk}^{\prime}(a, b, c)|
\label{eq:mae_3}
\end{equation}
where $m$ is the total length of the sequence and $\mathcal{S}$ stands for the set of 20 amino acids. During model selection (Supp.~Methods~\ref{appendix:model_selection}), we kept track of $MAE_{1}$ and $MAE_{2}$ between the statistics of the model generated set and that of the validation set of natural binders for each model (Methods~\ref{sec:dataset_preparation}). For the best-selected model M, we extended the examination to include $MAE_{3}$, verifying its effectiveness in capturing high-order beta strand statistics. As presented in Figs.~\ref{fig:freq_check}B, D, F, the generated data exhibited a strong correlation with validation data at single-site frequency ($MAE_1$=0.00093, $R^2$=0.99871, Pearson correlation coefficient=0.99935), two-point correlation ($MAE_2$=0.00013, $R^2$=0.90624, Pearson correlation coefficient=0.95197) and three-point correlation ($MAE_3$=0.00003, $R^2$=0.46089, Pearson correlation coefficient=0.67889). We performed an additional correlation analysis comparing generated data with training data, which produced similar results (Figs.~\ref{fig:freq_check}A, C, E). These outcomes imply that the chosen Model M has the capacity to generate samples that closely mirror the statistical properties of natural beta strands. It is worth noting that the training of Transformers does not rely on a moment-matching procedure that constrains the moments of the learnt distribution to match the empirical moments (as is the case for single-site and pairwise frequencies in protein sequence modelling methods like Direct Coupling Analysis \cite{morcos2011}). On the other hand, the Transformer learns probabilities of single sequence sites conditioned on the sequence context, which enables it to capture global correlations and hence to well reproduce high-order statistics \cite{sgarbossa2023generative}. 

\subsection{Validation on physicochemical properties}
\label{appendix:physicochemical_validation}

We evaluated five physicochemical properties of peptide sequences: 
\begin{enumerate}
\item Net charge ($Char$), which reflects the overall electric charge of a peptide. The net charge of each peptide was calculated by summing the charge of all amino acids in a peptide. 
\item Hydrophobicity ($Hydro$), which indicates the preference of a peptide for nonpolar environments based on its constituent amino acids' reluctance to interact with water. The hydrophobicity of each peptide was calculated by averaging the hydrophobicity (using the Kyte-Doolittle scale \cite{kyte1982simple}) of all amino acids in a peptide. 
\item Molecular weight ($MW$), which denotes the aggregate mass of a peptide's constituent atoms. The molecular weight of each peptide was calculated by averaging the molecular weight of all amino acids in a peptide. 
\item Isoelectric point ($IP$), which identifies the pH value at which a peptide exhibits a net charge of zero. The isoelectric point of each peptide was estimated using Bio.SeqUtils.IsoelectricPoint module in Biopython package \cite{cock2009biopython}. 
\item Aromaticity ($Arom$), which measures the proportion of aromatic residues within a peptide. The aromaticity of each peptide was calculated by the relative frequency of aromatic amino acids (Phenylalanine, Tryptophan, Tyrosine) in a peptide.
\end{enumerate}
During model selection (Supp.~Methods~\ref{appendix:model_selection}), we computed the cumulative distribution functions (CDF) of these physicochemical properties of generated binders, comparing them to those in the validation and random sets for each model (Methods~\ref{sec:dataset_preparation}, Fig.~\ref{fig:biophysical_properties}). We first calculated the Kolmogorov's D-statistics ($D_{Char}$, $D_{Hydro}$, $D_{MW}$, $D_{IP}$, $D_{Arom}$) for each respective property between the natural and random sets. The respective computed  D-statistics were 0.048427, 0.39769, 0.20989, 0.05868 and 0.10316, while the respective p-values were $6.35304 \times 10^{-110}$, 0 (less than representable positive number in python), 0, $2.96223 \times 10^{-161}$ and 0, demonstrating a clear distinction between natural and random binders. We then calculated the Kolmogorov's D-statistic for each respective property between natural and generated sets (see Table \ref{tab:model_metrics}). For the best selected model M, the computed D-statistics were 0.00338, 0.00300, 0.00332, 0.00443 and 0.00684, while the respective p-values were 0.571, 0.719, 0.592, 0.242 and 0.013. Four of the five tests can not be rejected at a 0.05 significant level, which implies no significant difference between these properties for generated and the natural binders. Based on these results, we can conclude that our selected model M accurately captures the physicochemical properties of beta strands, with a minor deviation in the case of aromaticity. 

\subsection{Embedding comparisons with protein language models} \label{appendix: embedding}

\begin{figure}[htbp]
\centering
\includegraphics[width=\linewidth]{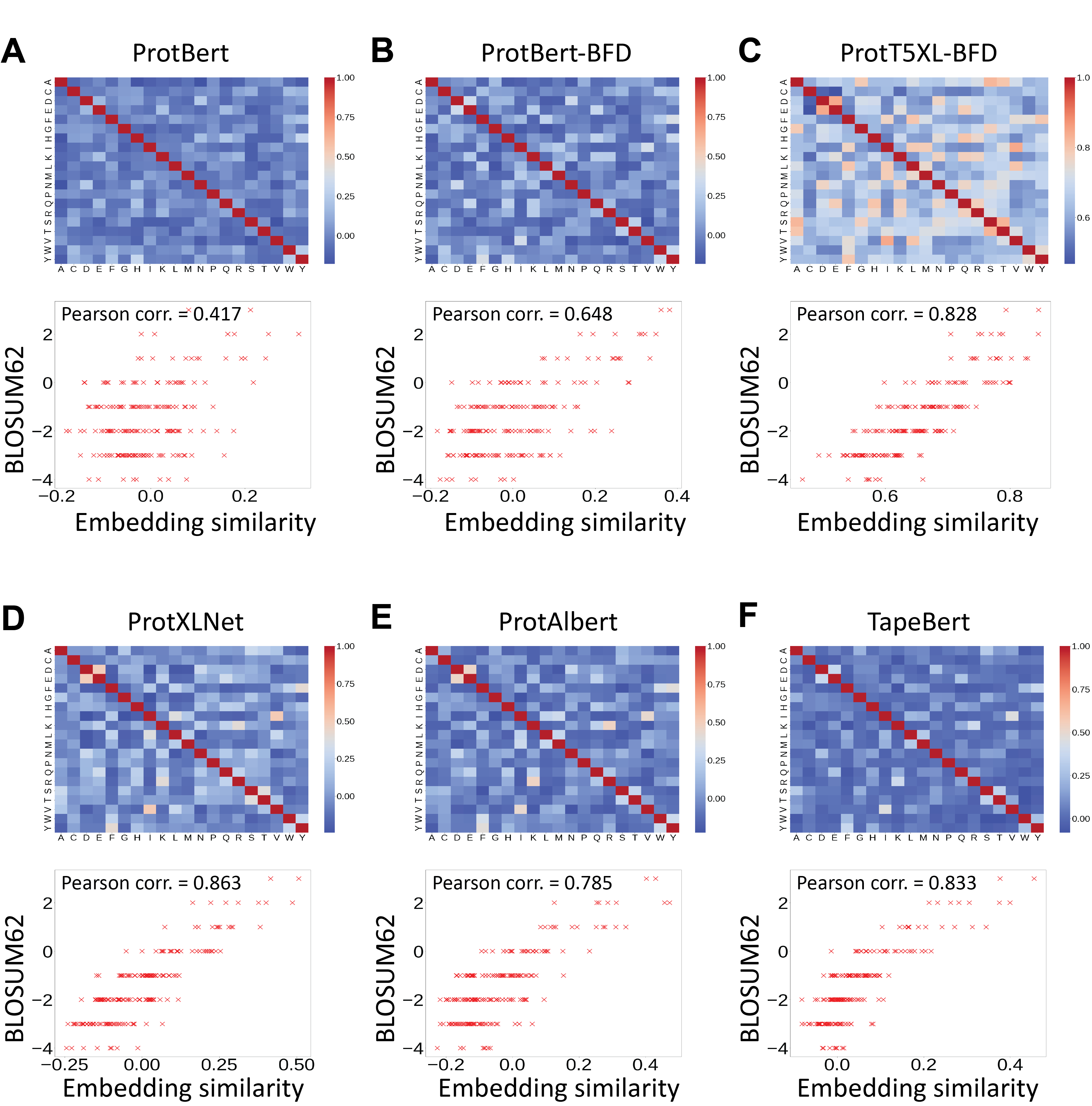}
\caption{\textbf{Correlation between cosine similarity of input embedding and BLOSUM62 substitution matrix scores across 6 pretrained protein language models} (see Table \ref{tab:protein_models_comparison}). (\textbf{A}) ProtBert: a BERT architecture model trained on UniRef100 dataset. (\textbf{B}) ProtBert-BFD: a BERT architecture model trained on BFD-100 dataset. (\textbf{C}) ProtT5XL-BFD: a T5 architecture model trained on BFD-100 dataset. (\textbf{D}) ProtT5XLNet: a XLNET architecture model trained on UniRef100 dataset. (\textbf{E}) ProtAlbert: a ALBERT architecture model trained on UniRef100 dataset. (\textbf{F}) TapeBert: a BERT architecture trained on Pfam dataset. }
\label{fig:embedding_analysis_benchmark}
\end{figure}

We calculated the embedding similarity between amino acids as in Equation~\ref{eq:embedding_similarity}, resulting in a 20 by 20 matrix. To avoid repeated values, we used the upper triangle of this matrix and the BLOSUM62 scores, then computed their Pearson correlation (Fig.~\ref{fig:embedding_analysis_benchmark}, Fig.~\ref{fig:embedding_analysis}B). 

By comparing the models' embeddings, we found that our model can learn amino acid representations comparable to protein language models with millions of parameters pretrained on millions of sequences, whose amino acid embedding similarity scores have a Pearson correlation with the BLOSUM62 scores ranging between $\sim0.42$ (for ProtBert) and $\sim0.86$ (for ProtXLNet), see Fig.~\ref{fig:embedding_analysis_benchmark}.

\subsection{Hardware}
\label{appendix: hardware}
We conducted database construction, data preparation, training and analysis work on the Imperial High Performance Computing cluster. Our model implementation was mainly built on the open-source Pytorch library \cite{paszke2019pytorch} and d2l package \cite{zhang2021dive}. We employed parallel computing \cite{li2020pytorch} to train the model (300K steps) on two RTX6000 GPUs over a duration of 50 hours.

\clearpage

\section{Supplementary Tables and Figures} 

\begin{table}[htbp]
\centering
\begin{tabular}{l|c c c c c c c}
\hline
Model & $N_{L}$ & $d_{model}$ & $d_{ff}$ & $h$ & $P_{drop}$ & $\epsilon_{ls}$ & train steps\\
\hline
base & 6 & 512 & 2048 & 8 & 0.1 & 0.1 & 100K \\
\hline
A & 6 & 512 & 2048 & \textcolor{red}{4} & 0.1 & 0.1 & 100K \\
B & 6 & 512 & 2048 & \textcolor{red}{16} & 0.1 & 0.1 & 100K \\
\hline
C & \textcolor{red}{2} & 512 & 2048 & 8 & 0.1 & 0.1 & 100K\\
D & \textcolor{red}{4} & 512 & 2048 & 8 & 0.1 & 0.1 & 100K\\
E & \textcolor{red}{8} & 512 & 2048 & 8 & 0.1 & 0.1 & 100K\\
\hline
F & 6 & \textcolor{red}{256} & 2048 & 8 & 0.1 & 0.1 & 100K\\
G & 6 & \textcolor{red}{768} & 2048 & 8 & 0.1 & 0.1 & 100K\\
\hline
H & 6 & 512 & 2048 & 8 & \textcolor{red}{0.0} & 0.1 & 100K\\
I & 6 & 512 & 2048 & 8 & \textcolor{red}{0.2} & 0.1 & 100K\\
\hline
J & 6 & 512 & 2048 & 8 & 0.1 & \textcolor{red}{0.0} & 100K\\
K & 6 & 512 & 2048 & 8 & 0.1 & \textcolor{red}{0.2} & 100k\\
\hline
L & 6 & 512 & 2048 & 8 & 0.1 & 0.1 & \textcolor{red}{300K}\\
M & 6 & 512 & 2048 & 8 & \textcolor{red}{0.2} & 0.1 & \textcolor{red}{300K}\\
\hline
\end{tabular}
\caption{\textbf{Transformer architecture hyperparameters scanned for model selection.} Deviations from the original Transformer architecture from \cite{Vaswani2017} (Model base) are emphasized in red.}
\label{tab:model_parameters}
\end{table}

\begin{table}[htbp]
\centering
\begin{tikzpicture}
\node (tabular) {
\begin{tabular}{l|c c c c c c c c}
\hline
Model & $\mathcal{L_V}$ & $MAE_{1}$ & $MAE_{2}$ & $D_{Char}$ & $D_{Hydro}$ & $D_{MW}$ & $D_{IP}$ & $D_{Arom}$\\
&  & ($\times 10^{-5}$)&($\times 10^{-5}$)& ($\times 10^{-5}$)& ($\times 10^{-5}$)& ($\times 10^{-5}$)& ($\times 10^{-5}$)& ($\times 10^{-5}$)\\
\hline
base & -12.26 & 109.97 & 13.73 & \textcolor{red}{370} & 1246 & 890 &\textcolor{red}{501}&\textcolor{red}{506}\\
\hline
A    & -12.20    & 121.06 & 13.50 & \textcolor{red}{388} & 1075 & 677 & \textcolor{red}{483}&\textcolor{red}{496}\\
B    & -12.32     & 117.68 & 13.62 & \textcolor{red}{451} & 1269 & 876 & 587&\textcolor{red}{533}\\
\hline
C    & -13.36     & 122.35 & 14.11 & \textcolor{red}{655} & 1617 & 759 &746&\textcolor{red}{320}\\
D    & -12.19     & 115.48 & 13.74 & \textcolor{red}{356} & 1484 & \textcolor{red}{552} &\textcolor{red}{407}&706\\
E    & -12.60     & 126.99 & 13.35 & 224 & 1239 & 760 &\textcolor{red}{454}&\textcolor{red}{533}\\
\hline
F    & -12.58     & 119.64 & 14.03 & \textcolor{red}{260} & 1561 & 944 &\textcolor{red}{385}&\textcolor{red}{257}\\
G    & -12.69     & 116.09 & \textcolor{blue}{13.20} & \textcolor{red}{231} & 837 & 638 & \textcolor{red}{398} &795\\
\hline
H    & -16.70     & 209.84 & 15.38 & \textcolor{red}{243} & 2045 & 1151 &810&\textcolor{red}{252}\\
I    & -11.98     & 95.20 & 13.51 & \textcolor{red}{312} & 665 & \textcolor{red}{320} &\textcolor{red}{380}&741\\
\hline
J    & -13.20     & 128.11 & 13.28 & \textcolor{red}{356} & 1318 & 925 & \textcolor{red}{423} & 1354\\
K    & -12.29     & 294.22 & 16.02 & 638 & 4090 & 2377 &866&629\\
\hline
L    & -12.65     & 111.71 & 13.34 & \textcolor{red}{396} & 934 & 648 & \textcolor{red}{581} & 679\\
M    & \textcolor{blue}{-11.95}     & \textcolor{blue}{92.52} & 13.31 & \textcolor{red}{338} & \textcolor{red}{300} & \textcolor{red}{332} & \textcolor{red}{443} & 684\\
\hline
\end{tabular}
};
\draw[red, dashed, thick, rounded corners] ([xshift=0.15cm, yshift=0.55cm]tabular.south west) rectangle ([xshift=-0.30cm, yshift=0.20cm]tabular.south east);
\end{tikzpicture}
\caption{\textbf{Transformer model selection}. Metrics for model selection are evaluated on the validation set (Methods~\ref{sec:dataset_preparation}). Best results according to log-likelihood, and quality of reproduction of statistical properties are emphasized in blue. We highlighted in red when the D-statistics  of the Kolmogorov-Smirnov test suggests that the null hypothesis ($H_0$) cannot be rejected at a 0.05 significance level. $H_0$ posits that the validation and generated samples distributions are identical, while the alternative hypothesis ($H_1$) asserts that they differ. The selected Model M is enclosed within a red dashed box.}
\label{tab:model_metrics}
\end{table}

\begin{table}[htbp]
\centering
\begin{tabular}{|l|c|c|c|c|}
\hline
\textbf{Model name} & \textbf{Architecture} & \textbf{Number of Parameters} & \textbf{Dataset} & \textbf{Dataset Size} \\ \hline
ProtBert & BERT \cite{devlin2018bert} & 420M & UniRef100 \cite{suzek2015uniref} & 216M \\ \hline
ProtBert-BFD & BERT & 420M & BFD-100 \cite{steinegger2018clustering} & 2.1B \\ \hline
Prott5XL-BFD & T5 \cite{raffel2020exploring} & 2.8B & BFD-100 & 2.1B \\ \hline
ProtXLNet & XLNET \cite{yang2019xlnet} & 409M & UniRef100 & 216M \\ \hline
ProtAlbert & ALBERT \cite{lan2019albert} & 224M & UniRef100 & 216M \\ \hline
TapeBert & BERT & 92M & Pfam & 31M \\ \hline
\end{tabular}
\caption{Summary of pretrained protein language models used for input embedding comparison.}
\label{tab:protein_models_comparison}
\end{table}

\clearpage

\end{document}